# Multicore magnetite nanoparticles prepared by glass crystallisation and their magnetic properties

Markus Büttner, Frank Schmidl, Paul Seidel

*Friedrich Schiller Universität Jena, Institut für Festkörperphysik, Helmholtzweg 5, 07743 Jena, Germany*

+49 (0)3641 9 47410

+49 (0)3641 9 47412

Markus.Buettner@uni-jena.de

Christian Worsch, Christian Rüssel

*Friedrich Schiller Universität Jena, Otto-Schott-Institut, Fraunhoferstrasse 6, 07743 Jena, Germany*

## Abstract

A potassium alumina borosilicate glass with the composition $13K_2O*13Al_2O_3*16B_2O_3*43SiO_2*15Fe_2O_{3-x}$ was melted using $Fe_2O_3$ as raw material. The melt was dumped from a Pt-crucible with a downpipe in water which resulted in the formation of phase separated droplets with a size of around 100 – 150 nm. In this droplets, magnetite crystals with a size of around 10-20 nm were observed. These magnetite nanoparticles with superparamagnetic behaviour are arranged to larger aggregates. This leads to a higher effective magnetic radius. According to magnetisation measurements the particles show hysteresis. The ratio of remanent vs. saturation magnetisation is not as high as it is necessary for uniaxial anisotropy. It is possible to elude the phase separations by cooking the pulverized glass in concentrated sodium hydroxide. Additional temperature dependent magnetorelaxometry (TMRX) measurements show in the distribution of the relaxation of magnetic moments over the course of temperature two peaks at 13 and 39 K. According to an interparticle distance smaller than 5 $d_C$ (the core diameter) could that be a result of strong magnetic interactions. Other magnetic relaxation processes also explain this measured effect.

## Introduction

Magnetite is a magnetic material which naturally occurs in numerous geological formations (Klein et al. 1985). But although known since many centuries, and although magnetite is a subject of scientific research since decades, more or less amazing properties are still not fully understood. The use of magnetite in magnetic nanoparticles (MNP) provides us a wide range of applications in industry, biology and medicine. For example stable colloidal liquids of MNP, called ferrofluids, are used in damping, bearing, sealing and lubrication, as ink or for magnetic fractionation. For that purpose, magnetic nanoparticles are coated by organic agents and dispersed in appropriate solutions (Harris et al 2003, Tartaj et al. 2003, Görnert et al. 2002, Odenbach et al. 2003, Suzuki et al. 1995).

Ferrofluids find various applications among which the separation processes e. g. of microorganisms from contaminated water is to be mentioned. They are also candidats for a large variety of biomedical applications (Jordan et al. 1996, Pankhurst et al. 2003, Tartaj et al. 2003). Here, the separation of cancer cells from human blood is utilised for cancer diagnostics (Hergt et al. 1998, To et al. 1992). Furthermore, magnetic nanoparticles can be used for hyperthermia; here magnetite particles are preferably incorporated into cancer cells (Andrä et al. 1998, Schwalbe et al. 2002). Then an alternating magnetic field is supplied which leads to a selective heating of the particles and hence of the cancer cells which results in selective destruction of cancer cells (Chan et al. 1993, Hilger et al. 2002).

The production of magnetic nanoparticles for pharmaceutical applications is mostly done by wet chemical (Tartaj et al. 2003, Wagner et al. 2004), especially precipitation reactions. Other preparation routes described in the literature are e.g. flame pyrolysis (Buyukhatipoglu and Clyne 2010). The draw back accociated with wet chemical methods is that usually a mixture of various iron oxides, such as magnetite, maghemite and hematite are obtained. If the magnetic nanoparticles are precipitated from aqueous solutions, the existence of several phases and the maximum crystallite size limits the magnetic moment of the particles.

In order to avoid a permanent interaction between magnetic particles, the residual magnetisation should be close to zero. Otherwise, the magnetic moment of a particular nanoparticle is limited by his size (Woltz et al 2006). Therefore, in order to increase the magnetic moment, many nanoparticles might be aggregated. If this is done in an ideal manner, the interaction of the crystals with each other can be avoided (Schaller et al. 2009). Such so-called multi-core nanoparticles were prepared with $SiO_2$ (Yang et al. 2005) or organic (Dutz et al. 2009) shells using wet chemical methods.

The glass crystallisation process enables to prepare magnetite as a pure phase with a comparatively narrow crystallite size distribution (Woltz et al. 2006). These magnetite particles show a larger magnetic moment than particles precipitated from aqueous solutions. During the course of the further preparation process, spontaneous crystallisation during cooling the melt has to be avoided. This requires a rapid cooling of the melt e. g. by using a twin roller technique. This procedure enables the preparation of amorphous samples with high iron oxide concentrations, usually in the range from 20 to 35 %. These samples are subsequently thermally annealed in order to achieve the crystallisation of magnetite. Subsequent to the crystallisation process, the residual glassy matrix is dissolved in diluted acetic acid and as residue, the magnetic nanoparticles are obtained. In some studies glasses with glass compositions of

$(24-y)Na_2O*yAl_2O_3*14B_2O_3*37SiO_2*25Fe_2O_3$ with y = 8, 12, 14, 16 mol% droplets with sizes in the 100–1000 nm range enriched in iron and containing crystalline magnetite with crystallite sizes in the 25–40 nm range in an amorphous matrix were found (Harizanova et al. 2010). The application of a twin roller technique, however, is not necessary using this glass compositions. These iron rich phase-separated droplets may capable for applications as multicore MNP.

Using potassium instead of sodium should close the miscibility gap which leads to smaller phase separated particles, which should be favourable for many MNP applications. In this paper, the preparation of a glass with the mol % composition $13K_2O*13Al_2O_3*16B_2O_3*43SiO_2*15Fe_2O_3$ is described. The resulting physical, notably the magnetic properties of the eluted phase-separated droplets were reported. The magnetic properties can be characterised by using commercial magnetometers. In this paper first the structure (electron microscopy) is presented. Then, EDX investigations and XRD measurements to determine the element composition and the average crystallite diameter are reported. Afterwards magnetic measurements and measurements of the magnetic relaxation signals between liquid helium and room temperature (temperature dependent magnetorelaxometry – TMRX) are described. In addition to the standard magnetic properties such as susceptibility and remnant magnetisation the TMRX method can provide beneficially information on the properties of the particles.

## Experimental Procedure

### Sample Preparation

A glass with the mol% composition 13 $K_2O$ * 13 $Al_2O_3$ * 16 $B_2O_3$ * 43 $SiO_2$ * 15 $Fe_2O_3$-x was melted under oxidising conditions using $K_2CO_3$, $Al(OH)_3$, $B(OH)_3$, $SiO_2$ and $Fe_2O_3$ as raw material. The glass was melted in 150g batches at a temperature of 1400°C kept for two hours using a Pt crucible in a superkanthal furnace. Then the temperature was increased to 1480°C kept for 20 min, and subsequently casted on a copper block. The glass was remelted in a Pt-crucible with a downpipe. This downpipe has a length of 20 mm and an inner diameter of 2 mm. For temperature control, a thermo-couple was joined to the outlet. Then an $Al_2O_3$ tube closed at one side was taken and a hole with a diameter of 5 mm was drilled in the bottom. The open side of the tube was equipped with a water-cooled flange. Then the crucible was placed in the tube in that manner that the downpipe was inserted into the bottom hole. Subsequently, the tube was placed in a vertical furnace. For isolation and fixation, refractory material was used. The open end was isolated with mineral wool. To prevent pollution of the furnace, another open $Al_2O_3$ tube was put around the crucible outlet. Below the furnace, a bucket with 8 L water was placed. The schematic experimental setup is shown in Fig. 1. The furnace was heated to 1510°C using a constant heating rate of 10 K/min. The difference between the temperature of the furnace and the outlet of the crucible was 16 K. All temperatures mentioned in the following refer to those measured at the downpipe. At 1380°C the first drops flew through the downpipe. At temperatures above 1460°C up to 1494°C, the drop frequency is approximately constant 11±2 $s^{-1}$. Coming into contact with water, the drops with a diameter of approximate 5 mm burst into pieces with sizes in the range of 1 to 2 mm. The use of this setup enables to frit the glass melt under reproducible conditions.
To elute the MNP from the glassy matrix, the glass samples were grinded and boiled 24 h in concentrated sodium hydroxide solution.

**Measurement methods**

The samples were studied with respect to phase separation and occurring crystals using transmission-electron-microscopy (TEM, *Hitachi H8100*) and scanning electron microscopy (SEM) *Zeiss DSM 940 A*. The element distribution in the glass matrix and the crystals was measured using TEM and energy dispersive X-ray spectroscopy (EDX). Powdered samples were investigated using X-ray diffraction (XRD *Siemens D5000*) in order to obtain information on type and size of occurring crystals. For these investigations Fe-K radiation with a wavelength of $\lambda= 1.936$ Å and Cu-K$\alpha$ radiation with a wavelength of $\lambda= 1.5406$ Å were used. Magnetic properties were studied with a SQUID magnetometer (*S600X* Cryogenics Ltd.). Using this device enables to record magnetisation curves as a function of temperature. From these curves information on the susceptibility, the saturation magnetisation and the coercive field were obtained. It is also possible to conclude on the anisotropy of magnetic particles. As an additional tool for the characterisation of magnetic particles the temperature dependent magnetorelaxation (TMRX) method was used. In principle, this method measures the relaxation of magnetic moments between room temperature and liquid helium temperature and was developed to detect the Néel relaxation signals over the studied temperature range. We use a helium anti-cryostat for measurements in the temperature range from 4.2 to 320 K. While the sample is self-cooled to liquid helium temperature, the sample is magnetised for 1 s under a flux density of 1 mT and after a dead time of 20 ms the magnetisation relaxation signal is acquired for 1 s. The data acquisition is achieved by an axial SQUID gradiometer which is connected to a data analysis system including a PC. For further details see (Romanus et al. 2007).

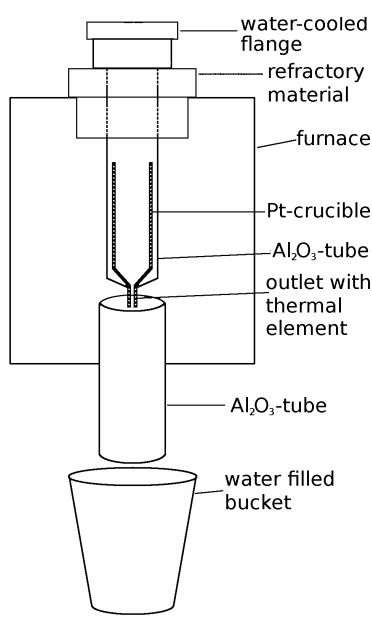

Fig. 1 Schematic of the experimental set up for melt quenching.

# Results

The melted and subsequently quenched sample had black coloration and is partially crystalline. The XRD-pattern of the sample is shown in Fig. 2. It exhibits comparable intense lines attributable to crystalline magnetite (JCPDS Nr. 19-0629) or maghemite (JCPDS Nr. 39-1346). Lines due to hematite, $Fe_2O_3$ (JCPDS Nr. 88-2359) are not observed. The lines are notably broadened and enable the calculation of a mean crystallite size of 20 nm using Scherrer equation (Klug and Alexander 1954).

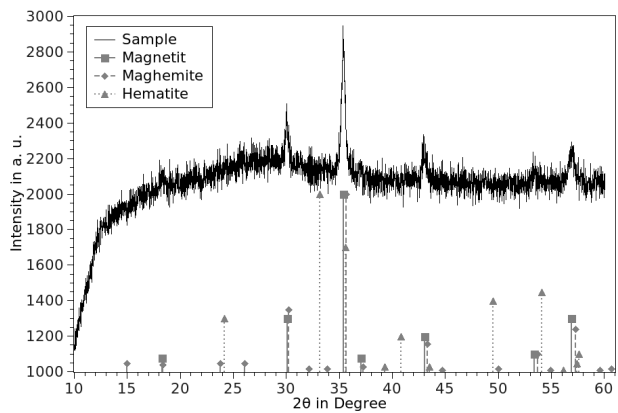

Fig. 2 XRD-pattern of the sample and the reflexes of magnetite, maghemite and hematite

A typical structure of the melted and subsequently quenched sample is shown in Fig. 3. The polished surface was etched for 20 s with 5% hydrochloric acid (HCl). The glass surface shows a fairly homogeneous structure with enclosed droplets with sizes in the range from 100 to 150 nm.

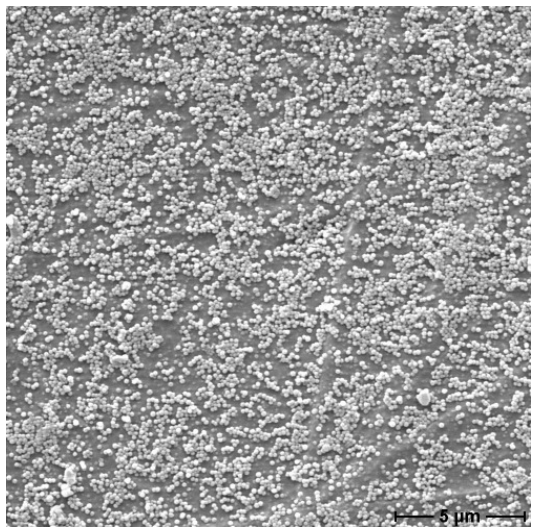

Fig. 3 SEM micrograph of a melted and quenched sample (etched in 5% HCl for 20 s)

In Fig. 4, TEM micrographs of the sample are shown. In analogy to Fig. 3, spherical structures are observed. Each droplet contains crystals with dark appearance which, according to the XRD-patterns, consist of magnetite (or maghemite). The sizes of these crystals are in the range from 10 to 30 nm, which is in good agreement with the crystallite sizes of 20 nm calculated from XRD-line broadening using Scherrer Equation.

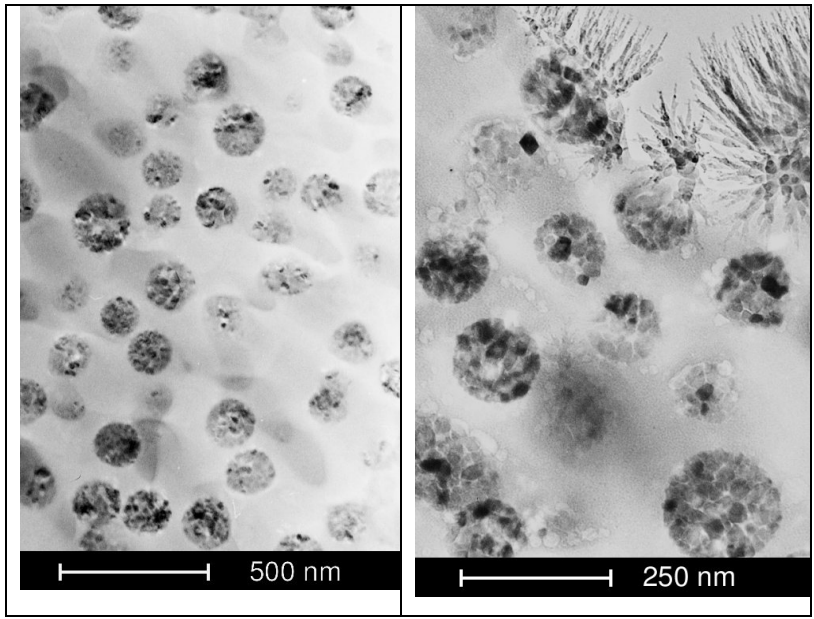

Fig. 4 TEM-micrographs of the sample (different magnifications)

Fig. 5 shows EDX-spectra recorded in the TEM. The spectra shown in Fig. 5 were collected in the glassy matrix (right) as well as in a phase-separated region (left). The peaks can be assigned to potassium, aluminium, silicon, iron and oxygen. By comparison, the glassy matrix contains higher potassium, aluminium and silicon concentrations than the droplets phase. The latter are enriched in iron.

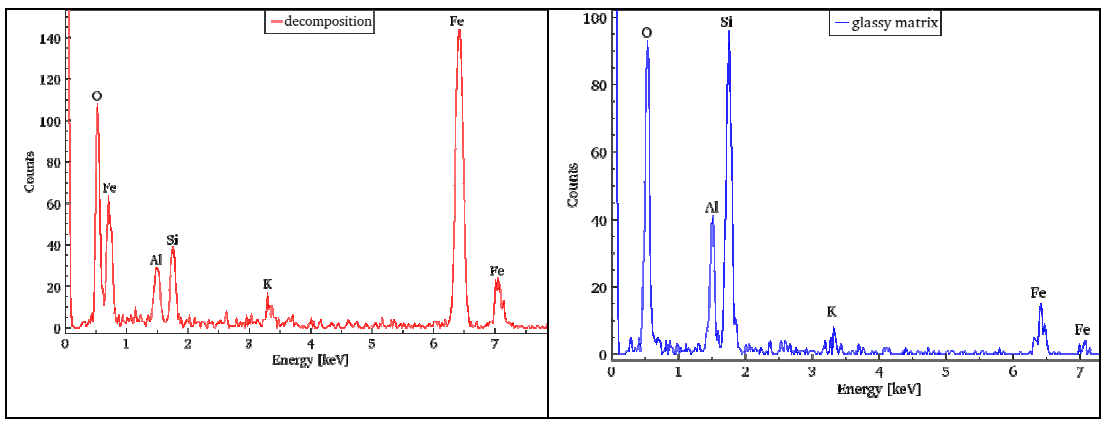

Fig. 5 EDX-spectra of the sample. Right: glassy matrix; left: droplets.

Fig. 6 shows TEM-micrographs of the glass powder which previously was treated with boiling concentrated sodium hydroxide solution for 24 h. This leads to the dissolution of the glass matrix. The remaining droplets are strongly agglomerated. The sizes of the droplets are in the range of 80 to 250 nm. Furthermore, some facetted structures are observed. The colour of the eluted particles after dissolution of the glass matrix is amber.

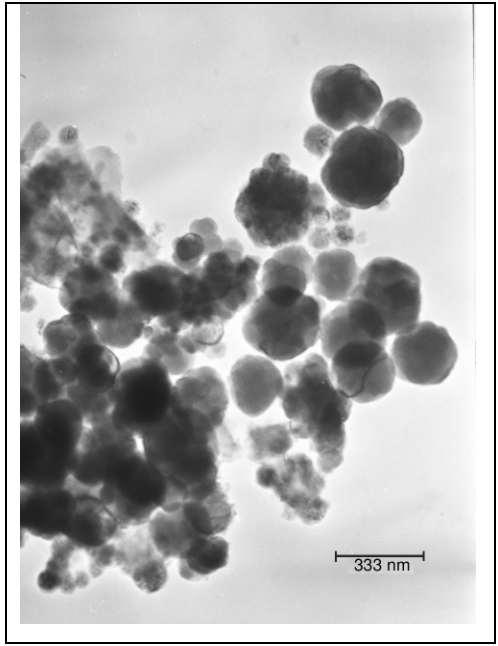

Fig. 6 TEM micrographs of the sample after dissolution of the glass matrix in boiling concentrated sodium hydroxide solution.

Since the magnetic properties of the produced particle system are the most important properties for possible applications the particles were magnetically characterised using a *S600X* SQUID magnetometer. All magnetic properties were measured using a magnetic flux density of 600 mT. Figure 7 shows two hysteresis loops recorded at temperatures of 5 and 306 K. From this hysteresis loops at different temperatures, the coercivity, the remanent magnetisation and the saturation magnetisation were obtained as a function of the temperature.

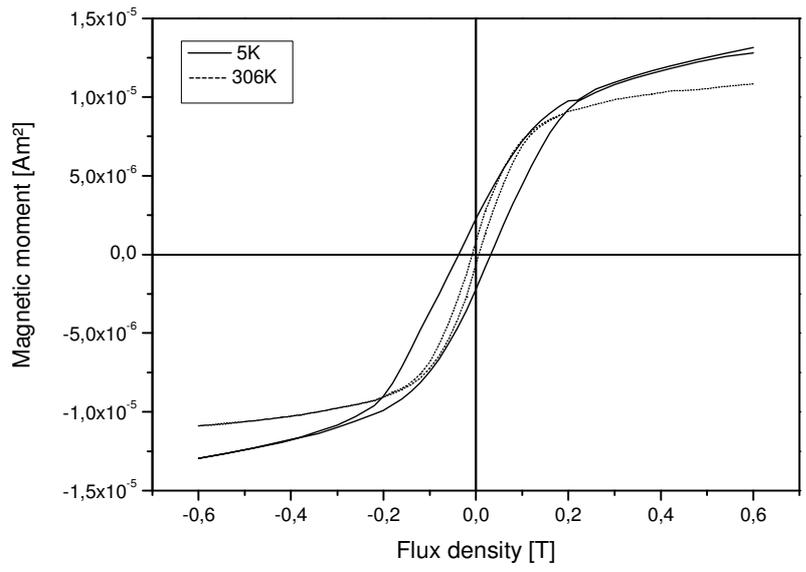

Fig. 7 Hysteresis loops at 5 K and 306 K

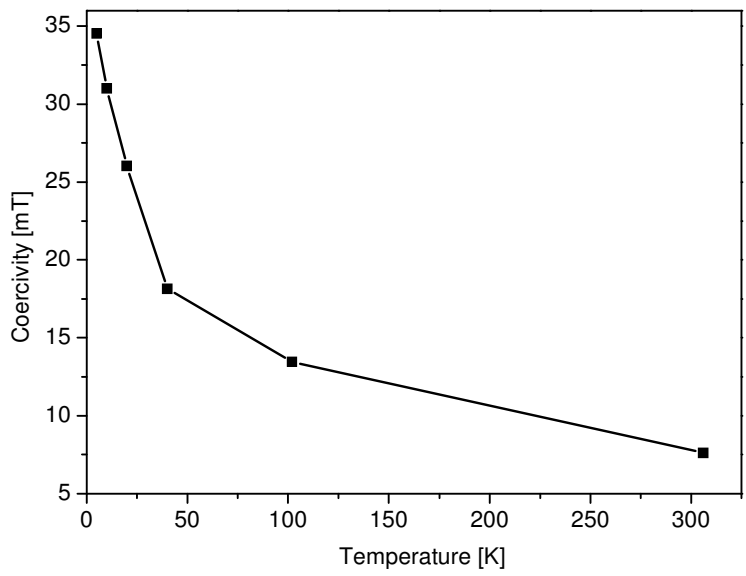

Fig. 8 Coercive filed over the course of temperature between 5 K and room temperature

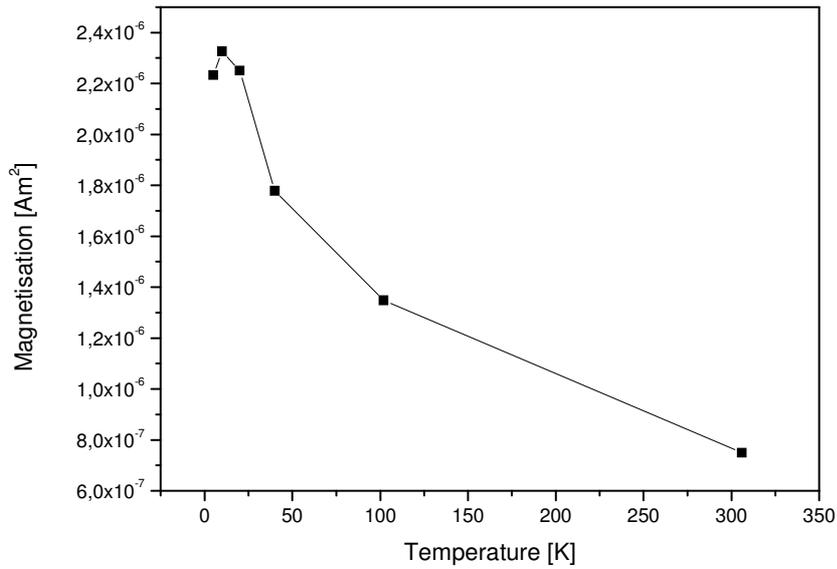

Fig. 9 Remanent moment over temperature between 5 K and room temperature

The coercivity (see Fig. 8) and the remanent magnetisation (see Fig. 9) both show the typical decrease with increasing temperatures. The values of the coercivity are similar to the values determined from magnetosomes (MNP consisting of pure magnetite produced by magnetotactic bacteria) (Büttner et al. 2011a) and MNP samples produced by precipitation from aqueous solutions (Büttner et al. 2011b) whereas the values for the remanent magnetisation are one order of magnitude lower. This is in good agreement with the objective target of this work.

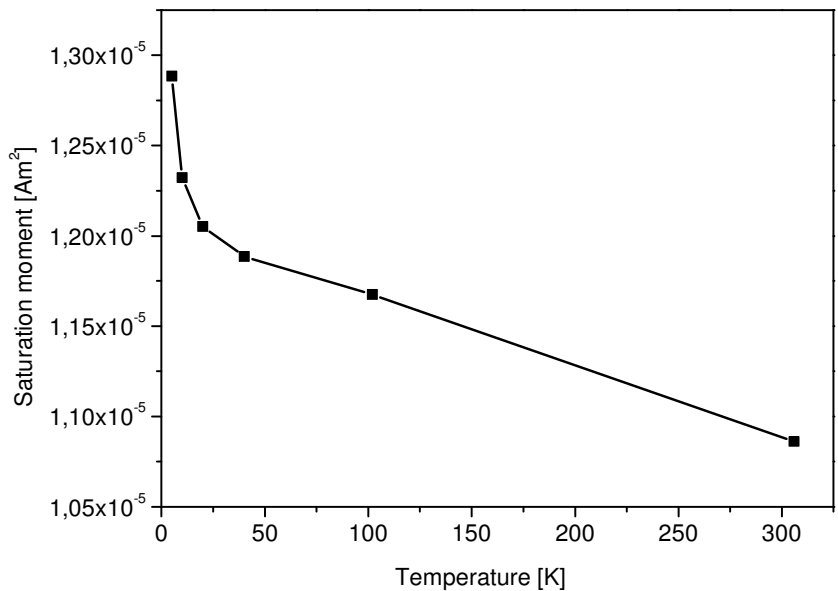

Fig. 10 Saturation moment over the course of temperature

In order to make a statement on the anisotropy of the particles, it is necessary to calculate the ratio of remanent magnetisation and saturation magnetisation at low temperatures. The value for the saturation magnetisation is taken from Fig. 10.

For a temperature of 5 K, a ratio of 0.17 is calculated which is notable smaller than the value of approximately 0.5 typical for uniaxial anisotropy. It can be concluded that the particles have a preferred direction of the resulting magnetisation. Figure 11 shows the result of the temperature dependent measurements of the magnetorelaxation signal. The measured values of the flux densities B(T) was multiplied with 293 K/T. Since the measured magnetorelaxation signals are a result of the numbers of particles N and their magnetically active volume V, the value NV was chosen as the scale for the y-axis in Fig. 11. Since SQUIDs do not supply absolute values, the units are arbitrary.

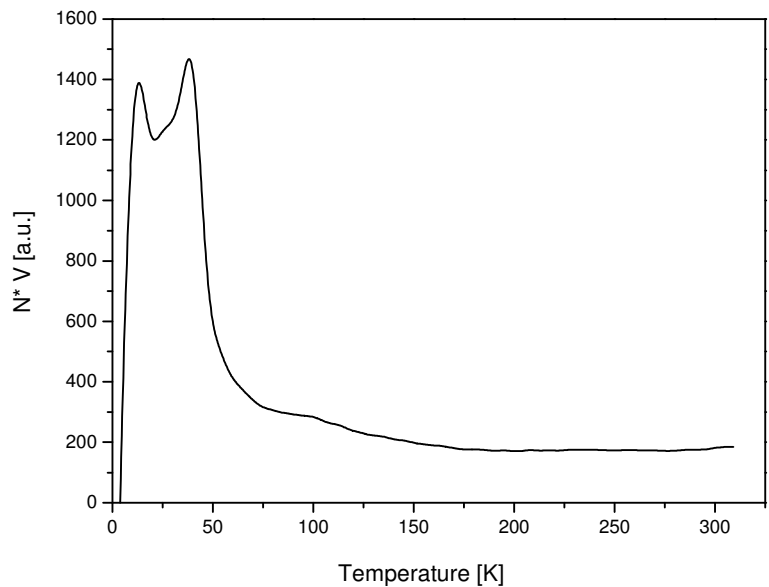

Fig. 11 N*V versus temperature curve calculated from the measured B(T) curve of the sample

The obtained curve in Fig. 11 shows a significant signal in the temperature range between 4.2 K and 60 K. This kind of signal, which is not a result of a Néel relaxation process, is already known from TMRX-measurements on other types of MNP (Büttner et al. 2011a, Büttner et al. 2011b).

## Discussion

The sample with the composition 13 K2O * 13 $Al_2O_3$ * 16 $B_2O_3$ * 43 $SiO_2$ * 15 $Fe_2O_{3-x}$, prepared by melting and subsequent quenching was partially crystalline. Using X-ray diffraction, the crystalline phase is identified as maghemite or magnetite. Especially hematite as crystalline phase was not observed. Maghemite is not a thermodynamically stable crystalline phase (under any conditions). Generally maghemite transforms irreversibly at T < TC to hematite and hence cannot be crystallised from a melt. This proves the occurrence of magnetite as the crystalline phase. Another hint is the colour of maghemite; it is red/brown, while the glass is completely black. The formation of magnetite by crystallising glass was previously observed by other authors (Romero and Rincon 1999; Woltz et al. 2004; Woltz et al. 2006; Harizanova et al. 2008, 2010, 2011; Wisniewski et al. 2011). In silicate glasses, in the most cases, dendritic growth of magnetite is observed. By contrast, in the present borosilicate glass composition, first droplet phase separation occurs. In the droplet phase, boron as well as iron is enriched, while silica is enriched in the matrix phase. The cubic magnetite crystals formed inside the droplets, possess a size of around 20 nm. Glass systems, containing sodium instead of potassium show similar characteristics (Harizanova et al. 2008, 2011). In these studies the glass was melted with $FeC_2O_4.2H_2O$ as raw material. For faster cooling, the reduced melts were drawn to fibres. However, the size of the phase separations is larger (100–1000 nm) and magnetic properties were not reviewed. Rapid quenching of the glass melt under reproducible conditions can control the size of the phase separated droplets and prevents dentritic growth of magnetite. With the used experimental setup and the experimental parameters supplied, the size of the phase separated droplets is 140 ± 40 nm. The exchange from sodium to potassium does not significantly affect the size of the crystallised magnetite crystals.

It is possible to dissolve the glassy matrix by using concentrated sodium hydroxide solution. The droplet phase was not dissolved and strongly agglomerated. The amber colour and the XRD-patterns indicate, that the $Fe^{2+}$ is partially oxidised and the composition of the iron oxide is shifted to hematite. The facetted structures in Fig. 3 might be residues of the glassy matrix or precipitated iron oxides, which were formerly dissoluted in the glass matrix. The latter could be responsible for the amber colour and the hematite reflex in the XRD-pattern.

In principle the magnetic behaviour of the particles is comparable with the properties of particles prepared by other production processes. The effects measured by TMRX in the temperature range between 4.2 K and 60 K might be caused by interactions of the magnetic moments (the so-called superspins) of the particles since some particles have only a small distance from each other. (see Fig.

4). On the other hand, Suzuki et al. (2009) and Sasaki et al. (2005) also investigated non-interacting particle systems with a superparamagnetic behaviour and reported ageing and memory effects at low temperatures using a FieldCooled-procedure. The dominating effect for the magnetorelaxation signals measured by TMRX cannot be labelled. Since the particles do not show uniaxial anisotropy a calculation of the magnetically active core sizes is not possible (Büttner et al. 2010). The basic magnetic properties such as coercivity and remanent magnetisation depend on the number of particles in a multiparticle core. When the numbers of particles in a multiparticle core and their sizes are reasonably comparable, the coercivity of the particles introduced in this work is two times higher than the coercivity of particles produced by precipitation reactions (Dutz et al. 2011). However, the size of the magnetite crystals produced via glass crystallisation is larger and the MNP contain more crystals.

The glass 13 $K_2O$ * 13 $Al_2O_3$ * 16 $B_2O_3$ * 43 $SiO_2$ * 15 $Fe_2O_{3-x}$ enables the production of MNP by simple melting and fritting of the melt. The MNP occurs in a glassy matrix, which allows the transport and storage without damage of the MNP. For applications, for example in hyperthermia, the MNP can be released from the glassy matrix.

## Conclusions

In a potassium alumina borosilicate glass with the composition $13K_2O*13Al_2O_3*16B_2O_3*43SiO_2*15Fe_2O_{3-x}$ the formation of phase separated droplets with a size of around 100 to 150 nm was observed. In this droplets, magnetite crystals with a size of around 10-20 nm occur. These magnetite nanoparticles are arranged to larger aggregates and show superparamagnetic behaviour. In magnetisation measurements the particles show hysteresis. The ratio of remanent vs. saturation magnetisation is not as high as in the case of uniaxial anisotropy. Additional temperature dependent magnetorelaxometry (TMRX) measurements show in the distribution of the relaxation of magnetic moments over the course of temperature two peaks at 13 and 39 K. According to an interparticle distance smaller than 5 $d_C$ (the core diameter) could that be a result of strong magnetic interactions. Other magnetic relaxation processes also explain this measured effect.

It is possible to separate the multicore magnetic particles by cooking the pulverized glass in concentrated sodium hydroxide.

*Acknowledgements*

The authors would like to thank S. Prass for the technical support, M. Röder for the magnetic measurements and T. Müller and M. Schiffler for helping us to finalise this paper. This work was supported by the EU project BIODIAGNOSTICS 017002.